# Analysis of Mandatory and Discretionary Lane Change Behaviors for Heavy Trucks

**Ding Zhao[1,*], Huei Peng[1], Kazutoshi Nobukawa[2],**
**Shan Bao[2], David J. LeBlanc[2] and Christopher S. Pan[3]**

**[1]Dept. of Mechanical Engineering, University of Michigan**
**[2]University of Michigan Transportation Research Institute**
**[3] Division of Safety Research, NIOSH, CDC**

* G041 Lay Auto Lab, University of Michigan
Ann Arbor, MI 48109-2133
Phone: 1-734-647-9732
Fax: 1-734-647-9732
E-mail: zhaoding@umich.edu

The behaviors of heavy vehicles drivers in mandatory and discretionary lane changes are analyzed in this paper. 640 mandatory and 2,035 discretionary lane change events were extracted from a naturalistic driving database. Variations in gap acceptance and lane change duration were investigated. Statistical analysis showed that mandatory lane changes are more aggressive in gap acceptance and lane change execution than discretionary lane changes. The results can be used for microscopic simulations, and design and evaluation of driver-assistant systems.

Topics / Driver Modeling, Active Safety

## 1. INTRODUCTION

Lane change, defined as a driving maneuver that moves a vehicle from one lane to another where both lanes have the same direction of travel [1] is a driving behavior that can depend on kinetics of multiple vehicles. Successful completion of a lane change requires attention to vehicles in both the original lane and the adjacent lane [2]. The fact that human could only look in one direction at a time, the presence of blind spot, and human perception limitations lead to anomalous behaviors [3], [4] or even crashes. In the US, in recent years between 240,000 and 610,000 reported lane-change crashes occur annually; resulting in at least 60,000 injuries [1].

Lane change behaviors have been actively studied for at least five decades. Early studies based on controlled experiments suffered from limitations including short test horizon and limited controlled settings [5]. The first large-scale naturalistic lane change study was conducted by the National Highway Traffic Safety Administration (NHTSA) in 2004. 500 lane changes with sedans and SUVs were used to examine lane change behaviors, including steering, turn signal and braking pedal usage, eye glance patterns, forward and rearward analysis and safety envelope [5]. Another NHTSA project, the 100-Car Naturalistic Driving Study, with primary goal to determine the main contributing factors associated with rear-end crashes, discuss the degree to which lane change events lead to rear-end conflicts [6]. The same database was further used to study the crashes and near-crashes lane change factors [1], [7].

While previous research provides preliminary conclusion about lane change for light vehicles, limited attention was paid to lane change behaviors for heavy vehicles [8]. According to the Federal Motor Carrier Safety Administration (2010), the number of heavy vehicles in the United States of America has increased by 70 % from 1975 to 2010. Meanwhile heavy vehicle-related incidents are consistently the leading cause of occupational fatalities, accounting for 13.6% of all work-related deaths[9], [10]. Truck drivers are five times more likely to die than the average worker on their duty [9], [10]. The unsafe actions of drivers are a contributing factor in about 70 % of the fatal crashes involving trucks [11]. As appealed by Occupational Safety and Health Administration, 'more public awareness of how to share the road safely with large trucks is needed' [11].

Lane Change Maneuvers (LCMs) are usually classified as either mandatory or discretionary (forced or unforced). When a lane change is required due to, for example, a lane drop or yielding to traffic near a ramp, it is called a Mandatory Lane Change (MLC). A lane change that is intended to improve the perceived driving conditions not urgently (e.g. passing a slow lead vehicle) is called a Discretionary Lane Change (DLC) [12]. Statistical analysis showed that there are obvious distinctions between these two kinds of maneuvers for sedans due to the difference in driving situations (forced or self-initiated) [13]. Two variables, gap acceptance and duration, are used as indicators for lane change decision and execution, with direct implication on lane change models and safety. Previous studies [14] on truck lane change used a bird's eye view video to study the surrounding traffic characteristics on lane change behaviors. However because of limited sampling size (only 28 truck lane changes), no stochastic distribution analysis was given. The results also had limitations with unknown driver information and mix-up with MLCs and DLCs which should be modeled separately.

## 2. NATRUALISTIC DRIVING DATABASE

The Integrated Vehicle-Based Safety Systems (IVBSS) database [15][16] is used in this research, which was collected in 2010 and maintained by the University of Michigan Transportation Research Institute (UMTRI). To collect the naturalistic data, 10 equipped Class 8 tractors were driven by 18 male commercial truck drivers from Con-way Freight for 10 months. The average age of the participants was 43 years (range = 28 to 63 years old) with an average of 13 years of driving experience. They were informed to drive naturally and assured that all the data collected would remain confidential and would be used only for safety research purposes (e.g., these data will not be shared with their employer).

Two types of routes (Fig. 1) were recorded: (a) pickup and delivery routes that involved delivering and picking up goods from local customers and (b) line-haul routes that involved moving the goods to one of several distribution terminals in the mid-western United States. Pickup and delivery routes ran only during the daytime, and long-haul routes ran only at night. Valid driving mileage consists of 601,844 miles of driving, or approximately 13,678 driving hours.

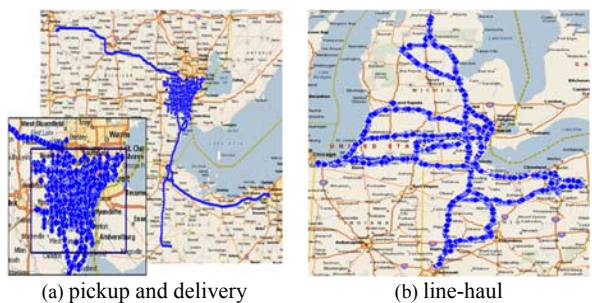

(a) pickup and delivery     (b) line-haul

Fig. 1. Recorded routes

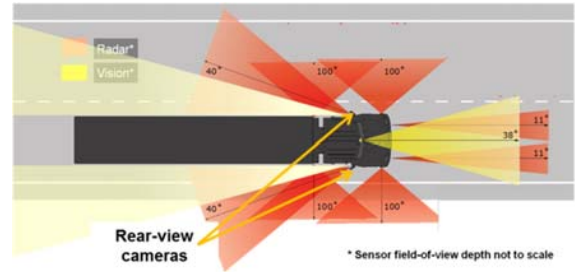

Fig. 2. Sensors equipped on commercial vehicle (not to scale)

Each truck was equipped with 8 radars and 3 exterior cameras (light yellow shade in Fig. 2 [17]). Over 500 channels regarding the driving environment, driver activity, system behavior, and vehicle kinematics were collected.

## 3. EVENTS EXTRACTION

198,132 lane change events were identified based on lane marker detection using front camera image. In this research, out of all the lane changes, 16,754 events with highway, high speed (55mph to 63mph), daytime (solar zenith angle between 0 and 96 degrees, or until civil dusk) were studied. To make the results more consistent, maneuvers to yield vehicles entering the freeway on the ramps from right side are selected to study MLC, and left lane changes to pass a slow lead vehicle are selected as a typical scenario for DLC. Both of the MLC and DLC scenarios are left lane change events. Events which are difficult to distinguish between these two types (e.g. lane changes happen when there are a merging vehicle from a ramp and a slow lead vehicle in front simultaneously) will not be studied in this paper. In the following contents, MLC and DLC are specifically used to refer to these two particular scenarios.

For MLC, the events were queried with an on-ramp appeared within 300m in front of starting point of lane change based on information from GPS and geographical location of on-ramp. DLC events were queried with two conditions: a) there was a slow lead vehicle before lane change (forward range $\in [1,100]$ m and forward range rate $\in [-10, 2.5]$ m/s) based on front radars; b) the lane change vehicle should pass one or more vehicle on the right within 60 sec after lane change based on side radars detection. Lane change events satisfy both MLC and DLC query conditions were eliminated to make the data more reliable. 640 MLC and 2,035 DLC were queried from the database. Manual checking was used to ensure the data quality.

## 4. LANE CHANGE GAP ANALYSIS

Gap acceptance refers to whether the adjacent gap in the target lane is safe to merge from the driver's perspective [18], which relates to various explanatory variables such as range, range rate, traffic conditions and host vehicle velocity. Fig. 3 gives an example scenario of a lane change situation. The front truck (Subject Vehicle or SV) is making a left lane change; while another sedan in the target lane (named as Principal Other Vehicle or POV). The (rear) lane change gap/range is defined as the distance between the rear edge of SV trailer and the front edge of the POV. In reality, the lane change decision is made a few seconds before the SV crosses the lane boundary. However, it is difficult to pinpoint the exact moment when SV driver makes the decision. To avoid ambiguity, in this study, the range and range rate are measured at the exact moment when SV crosses the lane boundary.

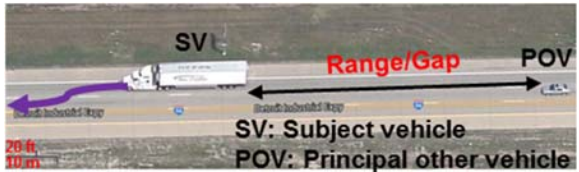

Fig. 3. Lane change scenario

### 4.1 Range and range rate estimation

For each lane change event, 10 frames before SV crossed the lane boundary are selected. First, the range in each frame is estimated based on image taken by the rear-view cameras mounted on the side mirror of the SV shown in Fig. 2. A pinhole camera model is used:

$$z_{c_i} = \frac{B_i}{b_i} z, \quad i = 1, 2, \dots, 10 \quad (1)$$

where $z_{c_i}$ is the distance between POV and the rearview camera, $B_i$ is the real size of a reference feature, $b_i$ is the size of the feature in the normalized image coordinates (i.e., $z = 1$) which is transformed from the original image coordinates on the camera retina in pixels using the camera parameters. The subscript $i$ represents that the measurements were extracted from frame $i$. It is found that the lane width at the front edge of POV (red line in Fig. 4 (b)) is a proper reference for range estimation. The real lane width of the was measured by the lane  Five points are extracted from the raw image manually shown in Fig. 4 (a), which indicate the lane boundaries and bottom line of the POV. After transforming these points into the normalized coordinates with no distortions and then by linearly extrapolating the two lane boundaries and horizontally extending the bottom line, the lane width on the image frame is determined. The extrapolated lines in the original image coordinates are shown in Fig. 4 (b). The Camera Calibration Toolbox for MATLAB® is used to obtain the intrinsic parameters (focal length, skewness, principal point, and distortion coefficient) for the coordinate transformations. This procedure only requires images of a checker board from multiple directions. For brevity, the standard camera calibration routine is omitted, which is explicitly explained in [19], [20]. The final range for a single frame is calculated as

$$R_i = z_{c_i} - l_t, \quad i = 1, 2, \dots, 10 \quad (2)$$

where $l_t$ is the longitudinal distance between camera and rear edge of the truck during trip $t$, which is calculated based on tractor geometry and the trailer type input by the truck driver at the beginning of the trip.

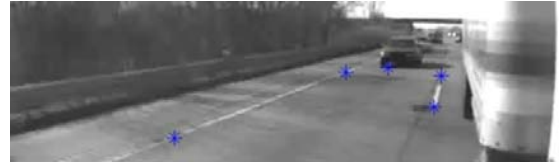

(a) Manual demarcation of feature points

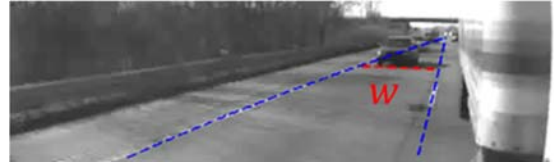

(b) Extrapolation of lane boundaries and POV bottom line

Fig. 4. Reference feature in the camera image frame

The range rate $\dot{R}$ is based on weighted least square, the weighting matrix $W$ is chosen as a diagonal matrix with the reciprocal of the estimated ranges as the diagonal terms. F is the vector of the frame number.

$$\dot{R} = (F^T W F)^{-1} F^T W R \quad (3)$$

where,
$$W = \begin{bmatrix} \frac{1}{R_1} & \cdots & 0 \\ \vdots & \ddots & \vdots \\ 0 & \cdots & \frac{1}{R_{10}} \end{bmatrix}, \quad \begin{array}{l} F = [1, 2, \dots, 10]^T \\[4pt] R = [R_1, R_2, \dots, R_{10}]^T \end{array}$$

It is noticed that in some frames, the lane marker is not clear. Those frames are dropped for data reliability. If the frame as the SV cross the lane boundary is dropped for this reason, linear extrapolation is used to estimate the lane change range. Events with fewer than 7 valid frames are considered invalid and will not be used in the following analysis. Events with POV with in Fast Approach Zone (FAZ) [1] (range $\leq$50 meters) are queried. 135 MLC and 156 DLC events with POV are finally identified.

### 4.2 Analysis of MLC and DLC gap acceptance

Statistical analysis is applied on range and range rate for MLC and DLC events.

Fig. 5 (a) shows that the range distribution of MLC and DLC. The normalized frequencies increase as range increases and saturate after range becomes large enough. It could be explained as it is more risky to make lane change with shorter range. However, when the range is large enough, the effect of range on lane change decision diminishes. In such case, lane change range mainly depends on traffic density.

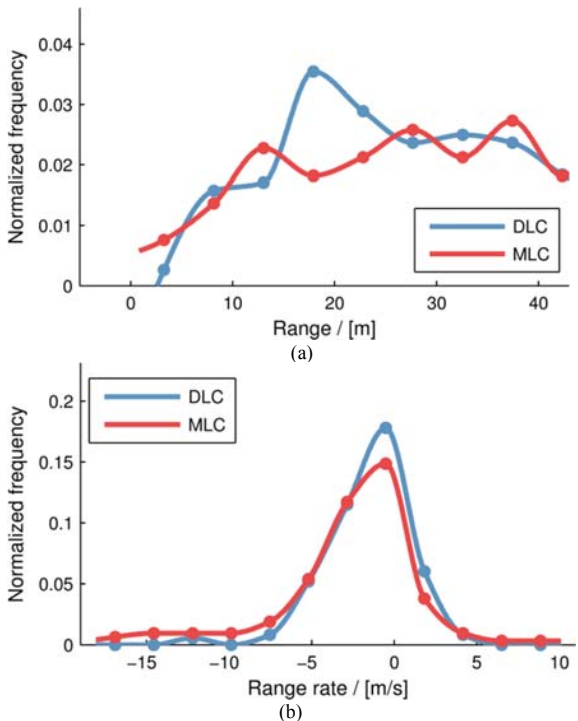

(a)

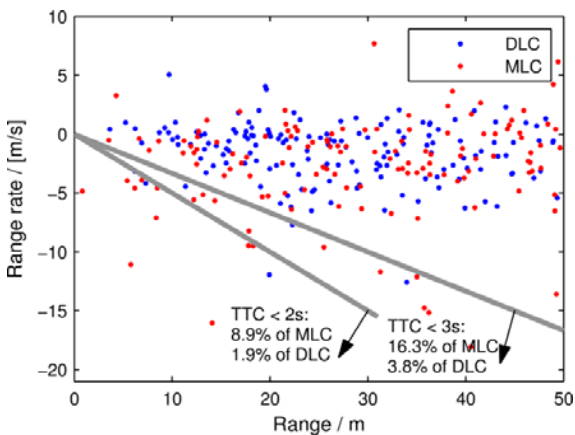

(b)

Fig. 5. Range (a) and range rate (b) histograms

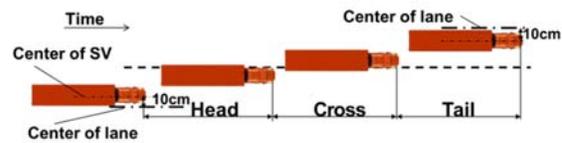

Fig.6. Range and range rate of all lane changes

Fig 5 (b) shows the range rate distributions of MLC and DLC. In both cases, the distributions follow bell-shaped curves. The range rate of MLC has a slight shift to the left and has a 'bigger tail' when the range rates are smaller than -5m/s.

Fig. 6 illustrates the relationship between range and range rate. The lower left corner represents more risky situations. The majority of MLC and DLC points are laid in the area with range rate between ±5m/s. Time To Collision (TTC), defined as the ratio of the range to range rate, is used to evaluate the risk of lane change. There are 8.9% of MLC with POV in FAZ with TTC less than 2 sec and 16.3% with TTC less than 3 sec. While only 1.9% and 3.8% of DLC have TTC less than 2 sec and 3 sec respectively. On average, MLC shows 4.5 times higher possibility than the cases of DLC. This indicates that MLC, as a forced maneuver, takes on more risk.

## 5. LANE CHANGE DURATION ANALYSIS

The lane change duration has been defined based on lane markers [21], steering angle [22], or heading angle [23]. In this research, we adopt the method based on lane markers. As shown in Fig. 7, the lane change procedure is divided into three stages. The head stage starts with a 10 cm offset from the center of the lane and ends when the SV first encroaches on the lane line between the original and destination lane. The cross stage follows until the vehicle completely crossed the lane boundary. The tail stage ends until SV reaches within 10cm offset in the destination lane. All 640 MLC and 2,035 DLC are used to study the lane change duration.

Fig. 7. The three stages of the lane changes

The distribution of lane change duration in each stage is shown in Fig. 8. Generalized Extreme Value (GEV) distribution is used to approximate the non-symmetric distribution (Fig. 8). GEV distribution is used as the statistic model because of its capability to model different shapes of tails, which has been used in other driver model behaviors such as [4]. It is shown that MLC and DLC generally follow GEV distribution in head portions, crossing portions and tail portions. The comparisons between the three periods are shown in Fig. 9. In head and tail portions, there is no significant difference between MLC and DLC. In crossing lane change portion, however, MLC has significantly shorter duration than DLC, which could be explained as SV drivers are under pressure and need to act quickly.

Mann-Whitney-Wilcoxon (MWW) test [24] is a non-parametric test of the null hypothesis that two populations are the same against an alternative hypothesis. Table 1 shows that the p-value for MWW tests between MLC and DLC in the three lane change stages. The crossing stage has a significantly small p value, representing that the probability to mistakenly reject the null hypothesis is less than 1%. The p value for head and tail stages are large (>30%). We could not distinguish between MLC and DLC in these two lane change stages statistically.

Table 1. MWW tests for lane change duration

|  | Head portion MLC vs DLC | Crossing portion MLC vs DLC | Tail portion MLC vs DLC |
|---|---|---|---|
| p-value | 0.4076 | <<0.01 | 0.3553 |

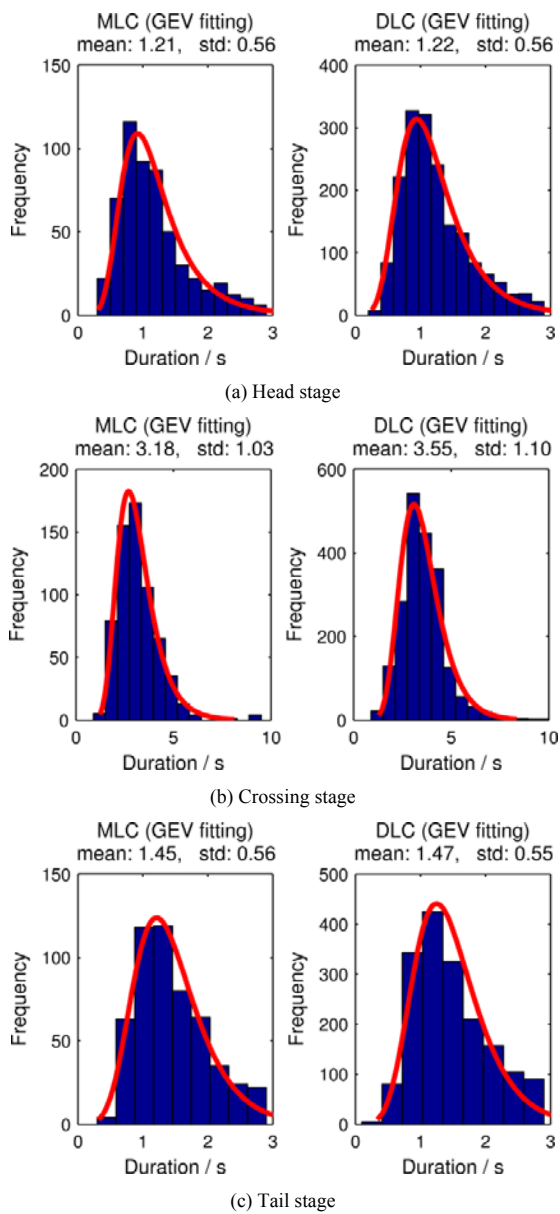

Fig. 8. Distributions of lane change duration

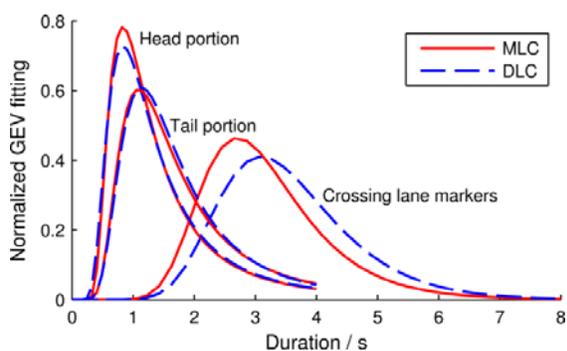

Fig. 9. Distributions of lane change duration

## 6. CONCLUSION

640 mandatory and 2035 discretionary lane change behaviors of heavy trucks were studied based on a naturalistic driving database. Distribution of range and range rate between SV and rear left POV were analyzed. Durations in three stages during lane changes were curve-fitted by generalized extreme value distribution and the difference between mandatory and discretionary lane change were proved with a Mann-Whitney-Wilcoxon test. Results show that mandatory lane changes are approximately 4.5 times more likely to have a critical gap condition than DLC and over 10% reduction of elapsed time in lane line crossing stage.

The results of this research could be used to build driver models for lane change decision and execution. The models not only should be able to emulate normal driving conditions but also generate risky driving scenarios. Such models may be used to evaluate lane change collision warning/avoidance algorithms for driver assistant systems. They could also be used to evaluate autonomous vehicle systems with a cut-in initiated by human driven vehicles.

## ACKNOWLEDGEMENT

The authors would like to acknowledge the generous sponsorship of the National Institute for Occupational Safety and Health (NIOSH). The authors also thank to Dr. Scott Bogard for the help on database query and Dr. Robert Goodsell for providing the software 'Viewer'.